\documentclass[11pt,a4paper]{article}

\usepackage{epsf}

\newcommand{\beq}{\begin{equation}}
\newcommand{\eeq}{\end{equation}}
\newcommand{\beqa}{\begin{eqnarray}}
\newcommand{\eeqa}{\end{eqnarray}}

\newcommand{\vs}{\vspace{-0.25cm}} 
\newcommand{\Mp}{M_{\pi^\pm}}
\newcommand{\MK}{M_{K^\pm}}
\newcommand{\Mn}{M_{\pi^0}}
\newcommand{\MKn}{M_{K^0}}

\begin{document}

\hfill {\tiny FZJ-IKP(TH)-2001-24}  
  
\bigskip\bigskip\bigskip 
 
\begin{center} 
{\Large\bf Isospin violation in low--energy\\[0.4em]
   charged pion--kaon scattering\footnote{Work supported in part
   by funds provided by the ``Studienstiftung des deutschen Volkes''.}
} 
\end{center} 
 
\vspace{.2in} 
 
\begin{center} 
{\large  Bastian Kubis\footnote{E-mail: b.kubis@fz-juelich.de},
 Ulf-G. Mei{\ss}ner\footnote{E-mail: u.meissner@fz-juelich.de}}

\bigskip 
 
\bigskip 
 
{\it Forschungszentrum J\"ulich,  
Institut f\"ur Kernphysik (Theorie)\\  
D--52425 J\"ulich, Germany}

\end{center} 
 
\vspace{.7in} 
 
\thispagestyle{empty}  
 
\begin{abstract} 
\noindent 
We evaluate the isospin breaking corrections 
to the scattering amplitude $\pi^- K^+ \to \pi^- K^+$ at threshold
in the framework of chiral perturbation theory. 
This channel is of particular interest
for the strong $2S-2P$ energy level shift
in pion--kaon bound states. While a prediction of this level shift is hampered
by a large uncertainty in the isoscalar scattering length, we find only a moderate
uncertainty of about 3\% in the electromagnetic corrections which are relevant
for the extraction of the scattering lengths from experiment.
\end{abstract} 
 
\vspace{1.3in} 
 
 
\centerline{Keywords:  
{\it pion--kaon scattering}, {\it electromagnetic corrections}, {\it chiral perturbation theory}}
 
\vfill

\newpage 
 

\noindent {\bf 1.}
Pion--kaon ($\pi K$) scattering near threshold is one of the cleanest processes to
test our understanding of chiral dynamics in the presence of strange quarks. 
It has been pointed out recently that the structure of the QCD vacuum might
change dramatically with increasing number of flavors, such that the scenario
for chiral symmetry breaking might be different for chiral SU(3) as compared to 
SU(2)~\cite{Descotes}: the quark condensate, which has been shown to be large
($B=|\langle 0|\bar{q}q|0\rangle| /F^2 \gg F$, where $F$ is the pion decay constant
in the chiral limit) for two flavors~\cite{CGL}, might be sizeably suppressed
in the three--flavor case.
$\pi K$ scattering might be used to test different scenarios for the SU(3) condensate
in much a similar way as $\pi\pi$ scattering has been used in~\cite{CGL} 
to show that the large condensate hypothesis indeed holds for two flavors.
As was pointed out in~\cite{genpiK}, one particular combination of the $\pi K$ S--wave
scattering lengths (the isoscalar one) is rather sensitive to deviations from the 
standard scenario.
The existing determinations of these
scattering lengths are, however, plagued by large uncertainties,
such that hope lies in the extraction of these parameters
from $\pi K$ bound states in the DIRAC experiment at 
CERN~\cite{DIRACpiK}. 
Measurements of both the partial width for the decay into the neutral channel,
$\Gamma_{\pi^0K^0} \propto |T^{\rm thr}_{\pi^-K^+ \to \pi^0K^0}|^2$,
and the strong $2S-2P$ energy level shift, 
$\Delta E_{2S-2P}^{\rm str} \propto T^{\rm thr}_{\pi^-K^+ \to \pi^-K^+}$,
allow to determine two independent combinations of the scattering lengths.

\medskip \noindent
In order to relate lifetime and strong energy level shift of $\pi^- K^+$ (or $\pi^+ K^-$) atoms to
particular combinations of the pion--kaon scattering lengths, one has
to make use of modified Deser formulae~\cite{Deser}
which include next--to--leading order effects in isospin breaking,
\beqa
\Gamma_{\pi^0K^0} &\propto& \left( a_0^{3/2} - a_0^{1/2} + \epsilon \right)^2 (1+\kappa) ~, 
\label{lifetime}\\
\Delta E_{2S-2P}^{\rm str} &\propto& \left( a_0^{3/2} + 2 a_0^{1/2} + \epsilon' \right) (1+\kappa') ~,
\label{shift}
\eeqa
(see~\cite{GGLR,pipatom} for equivalent formulae for the cases of pionium and pionic hydrogen). 
Here, $\epsilon(\epsilon')$ represents the isospin violating correction in the regular part of
the scattering amplitudes $\pi^- K^+ \to \pi^0 K^0 (\pi^- K^+)$ at threshold, while $\kappa(\kappa')$ 
is an additional contribution only calculable within the bound state formalism.
The correction $\epsilon$ to the scattering length relevant for the lifetime measurement
has been evaluated in~\cite{Kubis,Nehme}. Here, we complete that analysis by also calculating
the isospin violating shift $\epsilon'$ in the elastic charged channel.


\bigskip \noindent {\bf 2.}
The effective Lagrangian underlying our calculation
can be expanded at low energies according to
\beq
{\cal L}_{\rm eff} = {\cal L}^{(2)}
+ {\cal L}^{(4)} + \ldots ~,
\eeq
where the superscripts (2), (4) refer to the chiral dimension. 
Chiral power counting for the combined description
of the strong and the electromagnetic sector attributes the chiral dimension $p$
to all momenta involved, to the 
pseudo--Goldstone boson masses ($M_\pi$, $M_K$, $M_\eta$), as well as to 
the electric charge $e$. 
Therefore any term of the form $e^{2j} q^{2k} M_\pi^{l} M_K^{m} M_\eta^{n}$
with $2j+2k+l+m+n=4$ is counted as fourth order.

\medskip \noindent
The Lagrangian ${\cal L}^{(2)}$ leads to the well--known lowest--order mass formulae
for pions and kaons including leading electromagnetic effects,
\beqa
\Mp^2 = B(m_u+m_d)+2Ze^2F^2 ~, && \Mn^2  = B(m_u+m_d) + {\cal O}\left((m_u\!-\!m_d)^2\right) ~, \\
\MK^2 = B(m_u+m_s)+2Ze^2F^2 ~, && \MKn^2 = B(m_d+m_s) ~,
\eeqa
as well as to the Gell-Mann--Okubo relation valid at this accuracy, 
\beq
3M_\eta^2 = 2\MK^2 + 2\MKn^2 - 2\Mp^2 + \Mn^2 + {\cal O}\left((m_u\!-\!m_d)^2\right) ~.
\eeq
$m_{u,d,s}$ refer to the quark masses,
the low--energy constant $B$ is linked to quark condensate
as already mentioned above,
and $F$ is the common meson decay constant in the chiral limit.
The constant $Z$ accompanying the electromagnetic corrections to the charged meson masses
can be fixed from the leading order pion mass difference, $\Delta_\pi = \Mp^2-\Mn^2 = 2Ze^2F^2$,
as the strong pion mass shift of order $(m_u-m_d)^2$ is numerically tiny. 
Furthermore, the light quark mass difference induces the isospin breaking effect of $\pi^0 \eta$
mixing, where the mixing angle, at leading order, is given by
\beq
\epsilon = \frac{\sqrt{3}}{4}\frac{m_u-m_d}{m_s-\hat{m}} ~,
\eeq
where $\hat{m}=(m_u+m_d)/2$.
We will express all isospin breaking effects due to the light quark mass difference
in terms of~$\epsilon$.
In this work, the isospin symmetry limit is defined according to 
e.g.~\cite{GGLR}, i.e.\ using the \emph{charged} meson masses
$\Mp$, $\MK$ as a reference. Neutral meson masses entering via loop diagrams
will be expanded around this limit, to leading order in isospin breaking.
We therefore introduce a unified counting scheme for isospin violating effects,
combining both electromagnetic and strong effects in a common parameter 
$\delta \in \left\{e^2,\,m_u-m_d\right\}$.

\medskip \noindent
For the description of the process $\pi^- K^+ \to \pi^- K^+$ at one--loop level,
several low--energy constants of the fourth--order Lagrangian are needed, for
which we refer to the literature:
the Lagrangian involving the strong low--energy constants $L_i$ was defined 
in~\cite{NPB250}, while the operators corresponding to the electromagnetic
contants $K_i$ can be found in~\cite{Urech}.
For numerical evaluations, we use the central values and error estimates for the
hadronic low--energy constants as given in \cite{daphne}. For the
electromagnetic ones, we use the estimates obtained via resonance saturation
in \cite{BauUre}, and add error bars of natural size ($\pm 1/16\pi^2$)
uniformly.


\bigskip \noindent {\bf 3.}
Pion--kaon scattering in the isospin limit can be described by
two independent amplitudes $T^\pm$ which are isospin--even and --odd,
respectively,
\beq
T_{\alpha\beta} = \delta_{\alpha\beta} T^+ +\frac{1}{2} [\tau_\alpha,\tau_\beta] T^-
\eeq
($\alpha$, $\beta$ are the isospin indices of the pions), and which are related
to amplitudes of definite total isospin $1/2$ and $3/2$ by
$3 T^+ = T^{1/2} + 2T^{3/2}$, 
$3 T^- = T^{1/2} - T^{3/2}$.
These amplitudes (which are functions of the usual Mandelstam variables $s$, $t$, $u$
subject to the constraint $s+t+u=2(M_K^2+M_\pi^2)$)
can be decomposed into partial waves $t_l^I(s)$ according to
\beq
T^I(s,t,u)=16\pi \sum_l (2l+1) t_l^I(s) P_l(\cos\theta) ~,
\eeq
where $\theta$ denotes the scattering angle in the center--of--mass system,
and $P_l(\cos\theta)$ are the Legendre polynomials.
The real parts can be expanded at threshold in terms of scattering lengths 
($a_l^I$) and effective ranges ($b_l^I$),
\beq
{\rm Re}\,t_l^I(s) = \frac{\sqrt{s}}{2} \, q^{2l} 
   \Bigl\{ a_l^I + b_l^I q^2 + {\cal O} \left(q^4\right) \Bigr\} ~,
\label{def:thrpar}
\eeq
for center-of-mass momenta $q$. We will only be concerned with corrections to the
S--wave scattering length in this letter and therefore note that this quantity
is linked to the real part of the amplitude at threshold (i.e.\ for $q=0$) by
\beq
{\rm Re}\,T^I_{\rm thr} = 8\pi \left(M_K + M_\pi\right)\, a_0^I 
 + {\cal O}(q^2) ~. \label{threxp}
\eeq
The pion--kaon scattering lengths in the isospin limit have been evaluated up to one--loop order
in~\cite{BKMpiK1}, and analytic expressions for the complete scattering amplitudes 
can be found in~\cite{BKM}, however no analytic formulae for the scattering lengths
were written down. We therefore quote these for $a_0^\pm$ here:
\beqa
a_0^- \!&=&\! \frac{M_KM_\pi}{8\pi F_\pi^2(M_K+M_\pi)} \Biggl\{ 1 
\,+\,
\frac{M_\pi^2}{F_\pi^2} \,\Biggl[ \, 8L_5^r 
\label{a0minus} \\
\!&-&\!  \frac{1}{16\pi^2} 
\Biggl\{
\frac{8M_K^2-5M_\pi^2}{2(M_K^2 - M_\pi^2)}  \log \frac{M_\pi}{\lambda}
-\frac{23M_K^2}{9(M_K^2 - M_\pi^2)} \log \frac{M_K}{\lambda} 
+ \frac{28M_K^2 - 9M_\pi^2}{18(M_K^2 - M_\pi^2)}  \log \frac{M_\eta}{\lambda}
\nonumber\\ && 
\quad +\frac{4M_K}{9M_\pi}
\frac{\sqrt{(M_K - M_\pi)(2M_K + M_\pi)}}{M_K+M_\pi}
  \arctan \Biggl( \frac{2(M_K + M_\pi)}{M_K - 2M_\pi} \sqrt{\frac{M_K - M_\pi}{2M_K + M_\pi}} \Biggr)
\nonumber\\ && \quad 
-\frac{4M_K}{9M_\pi}
\frac{\sqrt{(M_K + M_\pi)(2M_K - M_\pi)}}{M_K-M_\pi} 
  \arctan \Biggl( \frac{2(M_K - M_\pi)}{M_K + 2M_\pi} \sqrt{\frac{M_K + M_\pi}{2M_K - M_\pi}} \Biggr)
\Biggr\} \Biggr] \Biggr\}
~, \nonumber \\
a_0^+ \!&=&\! \frac{M_K^2 M_\pi^2}{8\pi F_\pi^4(M_K+M_\pi)} \Biggl[
16\biggl( 2L_1^r + 2L_2^r + L_3 - 2L_4^r - \frac{L_5^r}{2} + 2L_6^r + L_8^r \biggr)  
\label{a0plus} \\ 
\!&+&\! \frac{1}{16\pi^2} \Biggl\{
\frac{11M_\pi^2}{2(M_K^2-M_\pi^2)} \log\frac{M_\pi}{\lambda}
-\frac{67M_K^2-8M_\pi^2}{9(M_K^2-M_\pi^2)} \log\frac{M_K}{\lambda}
+\frac{24M_K^2-5M_\pi^2}{18(M_K^2-M_\pi^2)} \log\frac{M_\eta}{\lambda} 
\nonumber\\ && 
\quad -\frac{4}{9}
\frac{\sqrt{(M_K - M_\pi)(2M_K + M_\pi)}}{M_K+M_\pi}
  \arctan \Biggl( \frac{2(M_K + M_\pi)}{M_K - 2M_\pi} \sqrt{\frac{M_K - M_\pi}{2M_K + M_\pi}} \Biggr)
\nonumber\\ && \quad 
-\frac{4}{9}
\frac{\sqrt{(M_K + M_\pi)(2M_K - M_\pi)}}{M_K-M_\pi} 
  \arctan \Biggl( \frac{2(M_K - M_\pi)}{M_K + 2M_\pi} \sqrt{\frac{M_K + M_\pi}{2M_K - M_\pi}} \Biggr)
+\frac{43}{9}
\Biggr\} \Biggr] ~. \nonumber
\eeqa 
The chiral analysis of these two quantities displays a remarkably different behavior
(which only becomes transparent when one discusses $a_0^\pm$ instead of $a_0^{1/2}$, $a_0^{3/2}$):
As was pointed out e.g.\ in~\cite{AB}, $a_0^-$ at order ${\cal O}(p^4)$ 
depends only on one single low--energy constant, $L_5$, which is in turn determined by
the ratio $F_K/F_\pi$, such that $a_0^-$ can be predicted to a very good accuracy,
$a_0^- = (0.0793 \pm 0.0006) M_\pi^{-1}$. On the contrary, no less than seven low--energy
constants enter the isoscalar scattering length $a_0^+$, some of them known to rather poor
accuracy, such that the chiral prediction for this quantity at one--loop order is plagued
by a very large uncertainty, $a_0^+ = (0.025 \pm 0.017) M_\pi^{-1}$.
Furthermore, the tree--level result for $a_0^-$ receives only a 12\% correction 
from one--loop contributions and counterterms (\emph{if}, as done here, one normalizes
the tree level result to $1/F_\pi^2$; 
see also the discussion in~\cite{Kubis}), 
while $a_0^+$ vanishes at tree level, and the contributions at order 
${\cal O}(p^4)$ are rather large.
In fact, if we regard the kaon as heavy and expand the 
expressions in eqs.~(\ref{a0minus}), (\ref{a0plus}) in powers of $M_\pi/M_K$, 
$a_0^-$ receives contributions of odd powers of the pion mass only, while
$a_0^+$ scales with even powers of $M_\pi$, such that one has symbolically:
\beqa
a_0^- &=& \frac{M_KM_\pi}{8\pi F_\pi^2(M_K+M_\pi)} \left\{
1+ \frac{M_\pi^2}{\Lambda_\chi^2} \left( c_0^- + c_2^- \frac{M_\pi^2}{M_K^2} 
+ c_4^- \frac{M_\pi^4}{M_K^4} + \ldots \right)
+ {\cal O}(\Lambda_\chi^{-4}) \right\} ~,
\\
a_0^+ &=& \frac{M_KM_\pi}{8\pi F_\pi^2(M_K+M_\pi)} \left\{
\frac{M_\pi M_K}{\Lambda_\chi^2} \left( c_0^+ + c_2^+ \frac{M_\pi^2}{M_K^2} 
+ c_4^+ \frac{M_\pi^4}{M_K^4} + \ldots \right)
+ {\cal O}(\Lambda_\chi^{-4}) \right\} ~,
\eeqa
where $\Lambda_\chi = 4\pi F_\pi$ is the chiral symmetry breaking scale.
This makes it obvious that the one--loop contributions to $a_0^-$ are suppressed by one power of
$M_\pi/M_K$ with respect to those to $a_0^+$.
This behavior is completely analogous to what one finds for pion--nucleon scattering 
lengths (see e.g.~\cite{BKMa-}).
We note that the scattering length of the charge exchange process,
entering the $\pi K$ atom lifetime formula eq.~(\ref{lifetime}), is given by
$a_0(\pi^-K^+\to\pi^0K^0) = -\sqrt{2}a_0^-$, while the sum of isoscalar and isovector
scattering length enter the strong energy level shift eq.~(\ref{shift}),
$a_0(\pi^-K^+\to\pi^-K^+) = a_0^- + a_0^+$. We therefore expect a rather different accuracy
in the predictions for these two quantities already from the purely strong contributions.

\begin{figure}[htb]
\vskip 0.5cm
\centerline{
\epsfysize=5.cm
\epsffile{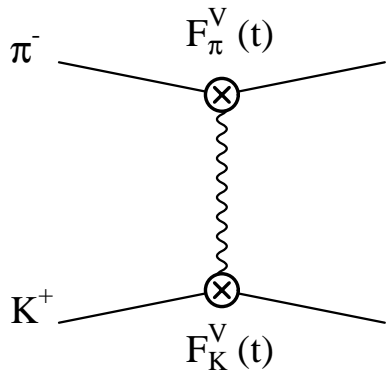}
}
\caption{One--photon exchange contributions to $\pi^- K^+ \to \pi^- K^+$.
\label{fig:1photex}
}
\end{figure}

\bigskip \noindent {\bf 4.} 
In the presence of isospin violation and in particular 
of virtual photon effects, the threshold expansion eq.~(\ref{threxp}) 
has to be modified: already at tree level, the one--photon exchange
contributing to $\pi^- K^+ \to \pi^- K^+$ cannot be decomposed into partial waves
and diverges at threshold. In order to allow for a direct matching to 
the description of the energy--level shift in the bound state formalism, 
we follow the prescription given in~\cite{pipatom} (for the analogous case of
$\pi^- p$ atoms) to circumvent these problems,
valid to leading order in the fine structure constant: the amplitude
$T_{\pi^- K^+ \to \pi^- K^+}$ can be uniquely decomposed into the piece
containing all diagrams that can be made disconnected by cutting one photon line,
and the remainder that contains at least one strong pion--kaon interaction vertex:
$T_{\pi^- K^+ \to \pi^- K^+} \,=\, T_{ex} + T_s$, where
\beq
T_{ex} \,=\, e^2 \left(\frac{u-s}{t}\right) F_\pi^V(t) F_K^V(t) ~.
\eeq
$T_{ex}$ is represented diagrammatically in fig.~\ref{fig:1photex}.
To the accuracy we are working,
the pion and kaon vector form factors are only needed in their isospin symmetric
representation to one--loop order as given in~\cite{GLff}.
Subtraction of the one--photon exchange diagram suffices to define the modified 
scattering length for $\pi^- K^+ \to \pi^- K^+$ in the presence of photons at tree level,
which is now taken to be $T_s$ at threshold instead of the full scattering amplitude.
The isospin violating corrections at tree level were already given in~\cite{Kubis}
(as for all other physical pion--kaon channels), they can be expressed 
in terms of the (electromagnetic) pion mass difference $\Delta_\pi$ as
\beq
a_0 \Bigl( \pi^- K^+  \to \pi^- K^+ \Bigr) \,=\,
  a_0^{\rm tree} \left\{1 + \frac{\Delta_\pi}{M_\pi M_K} \right\} 
  + {\cal O}(\delta^2)  
\,=\, a_0^{\rm tree} \Bigl\{1 + 0.018 \Bigr\} 
  + {\cal O}(\delta^2) ~, \label{a-+-+} 
\eeq
where, for convenience, we have used the lowest--order isospin symmetric scattering length
\beq
a_0^{\rm tree} = \left(a_0^- + a_0^+ \right)^{\rm tree} 
\,=\, \frac{M_KM_\pi}{8\pi F_\pi^2(M_K+M_\pi)} ~.
\eeq
No $m_u-m_d$ effects enter the scattering length at tree level.

\begin{figure}[htb]
\vskip 0.5cm
\centerline{
\epsfysize=3.5cm
\epsffile{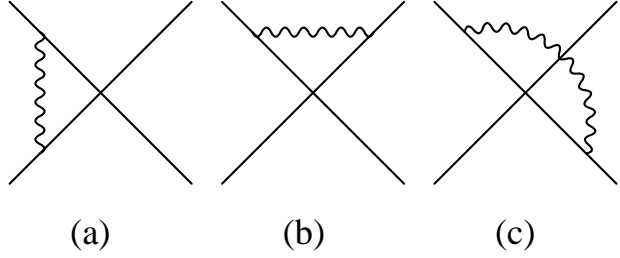}
}
\caption{Vertex corrections by photon loops contributing to $\pi^- K^+ \to \pi^- K^+$.
\label{fig:photloop}
}
\end{figure}

\medskip \noindent
However, even the scattering amplitude with the one--photon exchange diagrams subtracted
cannot be expanded at threshold according to eq.~(\ref{threxp}) 
once photon loop--corrections are included,
as the vertex correction diagrams, see fig.~\ref{fig:photloop}, 
induce a kinematical divergence, the Coulomb pole. 
The real part of the corresponding scalar loop function
\beq
G^{ab\gamma}(P^2) = -i \int \frac{d^4k}{(2\pi)^4}
\frac{1}{\Bigl( (q_a + k)^2 - M_a^2 \Bigr) 
         \Bigl( (q_b - k)^2 - M_b^2 \Bigr)
         \Bigl( k^2 - m_\gamma^2 \Bigr) } ~
\label{GpKg}
\eeq
(with $P = q_a + q_b$, and $a$ and $b$ referring to the particles involved), 
has the following threshold behavior in the $s$--channel (fig.~\ref{fig:photloop}(a)) 
for $s\approx(M_K+M_\pi)^2$:
\beqa
{\rm Re}\,\, G^{\pi K\gamma}(s) \!&=&\! - \frac{1}{32(M_K+M_\pi)\,q} \nonumber\\
&&\!+ \frac{1}{32\pi^2M_KM_\pi} \left\{ 2- \frac{M_K-M_\pi}{M_K+M_\pi} \log\frac{M_K}{M_\pi}
 - \log \frac{m_\gamma^2}{M_KM_\pi} \right\} 
+{\cal O}(q) ~,
\eeqa
while the corresponding loop functions in the $t$-- and $u$--channels 
(fig.~\ref{fig:photloop}(b) and (c), respectively) are regular at threshold,
\beqa
{\rm Re}\,\, G^{\pi K\gamma}(u) \!&=&\! 
\frac{1}{32\pi^2M_KM_\pi} \left\{  \frac{M_K+M_\pi}{M_K-M_\pi} \log\frac{M_K}{M_\pi} -2 
 + \log \frac{m_\gamma^2}{M_KM_\pi} \right\} 
+{\cal O}(q) ~,
\\
{\rm Re}\,\, G^{\pi\pi\gamma}(t) \!&=&\! \frac{1}{16\pi^2M_\pi^2}\log\frac{m_\gamma}{M_\pi} 
+{\cal O}(q) ~,
\eeqa
for $u=(M_K-M_\pi)^2$ and $t=0$, respectively. 
(For a full analytic expression for $G^{\pi K\gamma}(s)$, see~\cite{Kubis}.)
The infrared divergences, collected here
as terms involving $\log m_\gamma$, cancel at threshold. We therefore define the 
scattering length $a_0$ as the regular part of the amplitude $T_s$ at threshold,
\beq
{\rm Re} \,\, T_s \,=\, \frac{2\pi e^2 M_KM_\pi \, a_0^{\rm tree}}{q} 
 \,+\,  8\pi(M_K+M_\pi) \, a_0 \,+\,{\cal O}(q) ~.
\eeq


\bigskip \noindent {\bf 5.}
We now give the results for the
isospin violating effects in the scattering length for the physical channel 
$\pi^- K^+ \to \pi^- K^+$, 
which are expanded up to order $\epsilon$ and $e^2$.
We express these as relative corrections $\Delta a_0/a_0^{\rm tree}$.
The low--energy constants with the infinite part subtracted are
denoted by $L_i^r$, $K_i^r$, as done conventionally.
Note however that we do not display the dependence of these various
constants on the renormalization scale $\lambda$ explicitly.
All corrections given below are of course scale independent.

\medskip\noindent
The strong isospin violating contribution to the scattering length 
$a_0(\pi^- K^+ \to \pi^- K^+)$
can be expressed as a relative correction to the lowest--order
isospin symmetric result according to
\beqa
\frac{\Delta^{\rm str} a_0}{a_0^{\rm tree}} \!&=&\! 
\frac{\epsilon M_\pi}{8\sqrt{3}F_\pi^2}
\Biggl\{
\frac{6M_K^3-24M_K^2M_\pi+19M_KM_\pi^2-3M_\pi^3}{2(M_K^2-M_\pi^2)} \log\frac{M_\pi}{\lambda}
\\ \!&-&\!
\frac{64M_K(M_K-M_\pi)}{9(M_K+M_\pi)} \log\frac{M_K}{\lambda}
+ \frac{74M_K^3-40M_K^2M_\pi-43M_KM_\pi^2+27M_\pi^3}{18(M_K^2-M_\pi^2)} \log \frac{M_\eta}{\lambda}
\nonumber \\ \!&+&\! 
\frac{160M_K^3-180M_K^2M_\pi-43M_KM_\pi^2+45M_\pi^3}{9(4M_K^2-M_\pi^2)}  
\nonumber
\\ \!&+&\! \left.
\frac{2M_K(16M_K+5M_\pi)}{9(M_K+M_\pi)}\sqrt{\frac{M_K-M_\pi}{2M_K+M_\pi}}
\arctan \Biggl( \frac{2(M_K+M_\pi)}{M_K-2M_\pi}\sqrt{\frac{M_K-M_\pi}{2M_K+M_\pi}} \Biggr) \right\}
~. \nonumber
\eeqa
\noindent
In addition, the electromagnetic corrections can be written as
\beqa \!\!\!&& 
\!\!\!\!\!\!\!\!\!\!\!
\frac{\Delta^{\rm em} a_0}{a_0^{\rm tree}}  
\\ 
\!\!\!&=& \!\!
\frac{\Delta_\pi}{M_\pi M_K} 
\Biggl\{ 1 \,+\, 
\frac{4}{F_\pi^2} \biggl[
4\Bigl(M_K^2+M_\pi^2\Bigr)\Bigl(L_4^r-2L_6^r-L_8^r\Bigr) 
- \Bigl(M_K-M_\pi\Bigr)\Bigl(M_K+3M_\pi\Bigr) L_5^r \biggr]
\nonumber\\ \!\!\!&&\!\!  + \,
\frac{1}{16\pi^2F_\pi^2} \, \Biggl[
\frac{4M_\pi^3+2M_K^2M_\pi+11M_KM_\pi^2-M_\pi^3}{2(M_K^2-M_\pi^2)} \,M_\pi \log \frac{M_\pi}{\lambda}
\nonumber\\ \!\!\!&& \qquad
-\frac{9M_K^3+85M_K^2M_\pi-23M_KM_\pi^2+M_\pi^3}{9(M_K^2-M_\pi^2)} \,M_K \log\frac{M_K}{\lambda}
\nonumber\\ \!\!\!&& \qquad
-\frac{16M_K^3+8M_K^2M_\pi+36M_KM_\pi^2+7M_\pi^3}{18(M_K+M_\pi)} \log\frac{M_\eta}{\lambda}
\nonumber\\ \!\!\!&& \qquad
-\frac{8M_K^4-444M_K^3M_\pi-60M_K^2M_\pi^2+111M_KM_\pi^3+16M_\pi^4}{18(4M_K^2-M_\pi^2)}
\nonumber\\ \!\!\!&& \qquad
-\frac{M_KM_\pi(16M_K^2-17M_KM_\pi-8M_\pi^2)}{9(M_K+M_\pi)\sqrt{(M_K-M_\pi)(2M_K+M_\pi)}}
\arctan\Biggl( \frac{2(M_K+M_\pi)}{M_K-2M_\pi}\sqrt{\frac{M_K-M_\pi}{2M_K+M_\pi}}\Biggr) \Biggr]
\Biggr\}
\nonumber\\ 
\!\!\! &-& \!\!
e^2 \, \Biggl\{ \,
\frac{8}{3}K_1^r + 8\biggl(\frac{M_K}{M_\pi}+\frac{1}{3}+\frac{M_\pi}{M_K} \biggr) K_2^r
- 4\biggl(4+\frac{M_\pi}{M_K} \biggr) K_3^r 
- 2\biggl(4-\frac{M_\pi}{M_K} \biggr) K_4^r
\nonumber\\ \!\!\!\!&& \qquad
+ \frac{2}{3}\biggl(\frac{M_K}{M_\pi}+\frac{16}{3}+\frac{M_\pi}{M_K}\biggr) K_5^r
+ \frac{14}{3}\biggl(\frac{M_K}{M_\pi}-\frac{20}{21}+\frac{M_\pi}{M_K}\biggr) K_6^r
\nonumber\\ \!\!\!\!&& \qquad
- 8\biggl(\frac{M_K}{M_\pi}+\frac{M_\pi}{M_K}\biggr)K_8^r
- 12\biggl(\frac{M_K}{M_\pi}+\frac{M_\pi}{3M_K}\biggr) \Bigl(K_{10}^r+K_{11}^r \Bigr)
 \Biggr\} 
\nonumber\\ \!\!\!&+&\!\! 
\frac{3e^2}{8\pi^2} \, \Biggl\{
\frac{\bigl(5M_K-M_\pi\bigr) M_\pi}{M_K^2-M_\pi^2}
  \log \frac{M_\pi}{\lambda}
- \frac{\bigl( M_K^2-M_KM_\pi+4M_\pi^2 \bigr)M_K}{(M_K^2-M_\pi^2)M_\pi}
  \log\frac{M_K}{\lambda} 
+\frac{M_K-M_\pi}{6M_\pi} \Biggr\}
~. \nonumber
\eeqa
Inserting the values for the low--energy constants as given in 
\cite{daphne,BauUre}, we find the following corrections to the one--loop isospin symmetric
prediction:
\beq
a_0\left(\pi^-K^+\to\pi^-K^+\right) \,=\, \left(a_0^- + a_0^+\right) \biggl\{ \,
\underbrace{( 1 \pm 16.1\% )}_{{\cal O}(p^4)} 
\,+   \underbrace{\!\phantom{(} 1.2\% \phantom{)}\!}_{{\cal O}(e^2)} 
  -\, \underbrace{( 0.3 \pm 3.2 )\%}_{{\cal O}(p^2e^2)} 
\,+\!\!\!\!\!\!
      \underbrace{\!\phantom{(} 0.2\% \phantom{)}\!}_{{\cal O}(p^2(m_u-m_d))} \!\!\!\!\!\!
\biggr\} ~. \label{-+corr}
\eeq
Due to the appearance of the isoscalar scattering length, 
the one--loop representation for the isospin symmetric amplitude at threshold 
is afflicted by a rather large uncertainty (of about 16\%). 
Even if one includes isospin breaking corrections,
this uncertainty still dominates the combined error for a chiral \emph{prediction} 
of this quantity by far. If, however, the aim is to \emph{extract} the sum
of isovector and isoscalar scattering lengths from the experimental measurement
of the strong energy level shift, the corresponding uncertainty in the electromagnetic
corrections amounts to only 3.2\% and is therefore reasonable with respect to the
experimental accuracy that is expected. The central values for the various electromagnetic
low--energy constants lead to a total isospin breaking shift of 1.1\%. We remark that
the \emph{strong} isospin breaking is a pure loop effect and (at least with the 
normalization of the tree level amplitude as chosen here) free of any counterterms.

\medskip \noindent
For comparison, we also rewrite the result of \cite{Kubis} in this very form:
\beqa
&& \hspace{-1cm}
a_0\left(\pi^-K^+ \to \pi^0K^0\right) \,= 
\nonumber \\
&&  \quad\,\,
- \sqrt{2} \, a_0^- \biggl\{ \,
\underbrace{( 1 \pm 0.8\% )}_{{\cal O}(p^4)} 
\,+\, \underbrace{\!\phantom{(} 0.8\% \phantom{)}\!}_{{\cal O}(e^2)} 
\,+\!\! \underbrace{\!\phantom{(} 0.5\% \phantom{)}\!}_{{\cal O}(m_u-m_d)} 
\!\!-\, \underbrace{( 0.8 \pm 0.7 )\%}_{{\cal O}(p^2e^2)} 
\,+\, \underbrace{( 0.7 \pm 0.2 )\%}_{{\cal O}(p^2(m_u-m_d))} \,
\biggr\} ~.~~~ \label{00corr}
\eeqa
In contrast to eq.~(\ref{-+corr}), eq.~(\ref{00corr}) allows for a precise \emph{prediction}
of the charge exchange amplitude at threshold: compared to the isospin symmetric result,
it is increased by 1.3\% (for the central values of the low--energy constants involved),
while the combined uncertainty is enlarged only from 0.8\% to 1.1\%. 
We note that, in contrast to what was found for the one--loop representations of $a_0^\pm$,
there is no \emph{systematic} reason for the difference in sizes of the electromagnetic
uncertainties in the two channels at hand. As we attribute error bars $\pm1/16\pi^2$
uniformly to all electromagnetic constants $K_i$, the total uncertainty is dominated by
those counterterms which happen to appear with the largest numerical prefactors.
For the process $\pi^-K^+\to\pi^0K^0$, the dominant of such terms turns out to be
$\frac{8}{3}e^2 {M_K}/{M_\pi} (K_{10}+K_{11})$ (for the relative corrections
to the scattering length, at leading order in $M_\pi/M_K$), while the very same 
combination dominates in $\pi^-K^+\to\pi^-K^+$, just with a larger coefficient,
$12e^2 {M_K}/{M_\pi} (K_{10}+K_{11})$.


\bigskip \noindent {\bf 6.}
In this letter, we have completed the analysis of leading--order isospin breaking
corrections to the pion--kaon threshold amplitudes that enter the description
of pion--kaon bound state properties, up to one--loop order. 
While previous findings on the 
$\pi^-K^+\to\pi^0K^0$ amplitude allow for a rather precise prediction of the lifetime,
with isospin breaking corrections and the combined uncertainty of a size of the order of 1\%
(modulo the additional bound states corrections that still have to be calculated),
a prediction for the strong contribution to the $2S-2P$ energy level shift 
is hampered by the poor knowledge of the (strong) isoscalar scattering length. 
The uncertainty for the \emph{extraction} of the linear combination $a_0^-+a_0^+$
as induced by electromagnetic corrections, however, is only about 3\% and therefore
sufficiently well under control. The isospin breaking shift for the central values
of the low--energy constants involved is again about 1\%. 
In order to complete an effective field theory analysis of pion--kaon atoms,
these results now have to be incorporated into a study along the lines
of \cite{GGLR} (for pionium) and \cite{pipatom} (for pionic hydrogen).

\bigskip \noindent {\it Note added:} During the preparation of this letter, 
a similar analysis of charged pion--kaon scattering
with essentially similar results appeared \cite{Nehme2}. 
We only note that a different convention was used to subtract the long--range
photon exchange: in \cite{Nehme2}, only the \emph{tree--level} exchange
was disregarded. The difference of the two approaches in the amplitude at threshold can be easily
expressed in terms of the radii of the pion and kaon vector form factors,
\[
T^{\rm thr}_{\rm Nehme} = T^{\rm thr}_{\rm KM}
- \frac{2}{3}e^2 M_\pi M_K \left( \langle r^2 \rangle_\pi^V + \langle r^2 \rangle_K^V\right) ~. 
\]

\end{document}